# A Revisit on Blockchain-based Smart Contract Technology

Fengkie Junis, Faisal Malik Widya Prasetya, Farouq Ibrahim Lubay, Anny Kartika Sari
Department of Computer Science and Electronics
Faculty of Mathematics and Natural Sciences
Universitas Gadjah Mada
Yogyakarta, Indonesia
{fengkie.junis, faisal.malik.w, farouq.ibrahim.l}@mail.ugm.ac.id, a_kartikasari@ugm.ac.id

*Abstract* — Blockchain-based smart contract has become a growing field in the blockchain technology. What was once a technology used to solve digital transaction issues turns out to have some wider usage, including smart contract. The development of smart contract can be traced from the numerous platforms facilitating it, however the issue on how well each platform works as oppose to each other has yet been fully explored. The usage of smart contract can be seen from the applications that are built on top of the smart contract platform, such as the tokenization of real world to virtual world assets. However smart contract contains several issues concerning security and codifying which could be solved by various tools that are proposed by existing research. This paper aims to revisit the blockchain-based smart contract technology in order to understand and discuss the research gaps gathered from existing research and to provide guidance for future research.

## I. INTRODUCTION

Centralized form of transaction, which are usually conducted with a trusted third party, provides various problems, such as a single point of failure. This could be solved by using blockchain technology, which provides a peer-to-peer transaction without the need of a third party [1]. In 2009, the release of Bitcoin, a decentralized cryptocurrency, has gathered interest in the blockchain technology field. The blockchain technology that used to be applied only for bitcoin peer-to-peer transaction has been also usable for other purposes, such as smart contract, which is a program that self-enforce the contract clauses on the blockchain.

Development relating to blockchain-based smart contract has been accumulating over the years. This development ranges from various platforms that facilitates blockchain-based smart contract, applications that utilized blockchain-based smart contract, tools in developing blockchain-based smart contract applications, and new proposals to improve the technology. However the problem is that the development has been accumulated over the years, and there is a lack of up-to-date review on the research and proposals that have been issued for the blockchain-based smart contract.

The fast paced development on this topic needs to be compiled, in order to identify the latest development of the technology, current best practices, and to identify the research gaps that needs to be addressed in future research. Therefore, in this paper we revisit recent articles to provide the readers with up-to-date review on the current situation of the blockchain-based smart contract technology.

This paper is strcutured as follows. In Section II, the general motivation and description of blockchain technology is presented. It is followed by the review on blockchain-based smart contract platforms in Section III. Section IV discusses the smart-contract based applications. In Section V, tools, proposals, and best practices are elaborated. Section VI contains discussion and a proposal of possible implementation of blockchain-based smart contract in academic field. Conclusion is presented in Section VII.

## II. GENERAL IDEA AND MOTIVATION BEHIND THE BLOCKCHAIN TECHNOLOGY

In order to understand blockchain, exploring the motivation behind its creation is essential to understand its significance. It was initially referred to as "chain of blocks" in Nakamoto's paper [1]. This technology enables digital cash to be spent directly from one party to another without any involvement of a third party. In broad sense, blockchain is an immutable, append-only linked list of blocks that are chained together in such way that the alteration of a block will alter its consecutive blocks. Fig. 1 illustrates a blockchain and how it works. Each block in a blockchain contains transactions, then it is hashed alongside previous block's hash. In this way, the hashes are linked together, rendering any alteration of previous blocks also need to change the next block.

One of the reasons behind the idea of blockchain invention was to avoid double spending, a problem in digital cash scheme that has been identified since the idea of digital cash was introduced by David Chaum in 1983 [2]. Double spending problem is a flaw where digital cash can be spent twice due to the nature of digital data being easily duplicated. There are two solution for double spending problem. The first one is centralized solution, by appointing a trusted third party to keep the accounting at the expense of a risk of single point of failure. Or, it can be done in decentralized peer-to-peer fashion to eliminate the risk of single point of failure with the risk of sybil-attack, an attack where some actor in the network can gain control by creating fake identities [3]. In the proposal paper of Bitcoin, Nakamoto proposed blockchain as a solution for double spending problem by creating a cryptographically-secured immutable digital ledger that is stored in a peer-to-peer manner by all involving party. Because Bitcoin aimed to be a decentralized peer-to-peer digital cash, it has to avoid double spending problem and to consider the possibility of sybil-attack to its network. The network is secured from the sybil-attack by using a consensus protocol based on–albeit, not limited to– proof-of-work algorithm to make the attack very expensive to be performed [1].

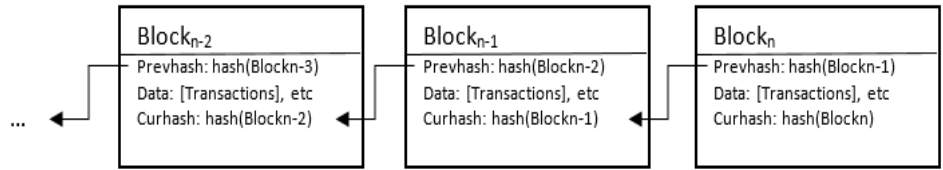

Fig. 1. Illustration of a blockchain

It turns out that the very same technology that underlies Bitcoin as a peer-to-peer electronic money, the blockchain, can be applied to many other fields such as financial services, healthcare, business and industry, IoT, and legal services. Specifically, several attempts has been made on several cases such as public notary, right management, document's proof-of-existence, authentication protocol, storage, anti-counterfeit, and internet applications [4, 5]. Therefore the research on blockchain technology gains interests from researcher of different fields.

### III. BLOCKCHAIN-BASED SMART CONTRACT

The notion of smart contract has been in existence since 1996. It was proposed by Szabo [6], which stated: *"A smart contract is a set of promises, specified in digital form, including protocols within which the parties perform on these promises."* [6]. There is a need to implement a mechanism to enforce contract execution. Like the solution for double spending problem, it can be implemented by two approaches: centralized approach where the control of contract execution is put on a trusted third party, or decentralized approach as in Bitcoin. The enforcement of smart contract can be viewed as a more generalized form of the double spending problem. Therefore, the concept of blockchain can be applied as a solution to this problem, this eliminate the risk of single point of failure in contract execution and enforcement.

In some sense, blockchain-based smart contract can be thought as a computer program that is executed on top of a blockchain network [7]. There are several platforms facilitating this, including Bitcoin, as it provides some limited non-Turing complete scripting capabilities, although it is not originally intended as a general smart contract platform. As identified by Bartoletti and Pompianu in early 2017, several smart contract platforms are Bitcoin, Ethereum, Counterparty, Stellar, and Lisk [9]. There are other platforms such as Cardano, Neo, Rootstock, and several others [8]. We will discuss these prominent smart contract platforms in this section.

Bitcoin was the first application to implement blockchain to facilitate transfer of monetary value in a peer-to-peer fashion over the internet. It facilitates some limited non-Turing complete scripting execution on the network but limited to only supporting several basic arithmetic, logical, and cryptographic functions such as hashing and digital signature verification. The limited expressiveness is due to this limited scripting capabilities and only small number of nodes that can process complex scripts [9].

The second platform is Ethereum. Introduced as a generalized programmable blockchain, it facilitates a Turing complete programming language to be executed on the network. However, the execution of a Turing complete programming language is risky due to the halting problem, where a program cannot be determined whether it will halt or not at some point in the future. Ethereum avoids this problem by introducing the concept of gas, a fundamental cost unit for the network, which is paid by a cryptocurrency called Ether to execute an Ethereum smart contract [9, 10]. An Ethereum blockchain can be thought as a transaction-based state machine with its own storage, memory, and a set of bytecode instructions, where a valid state transition consisting of transactions recorded in a block of the blockchain. This state machine is referred to as Ethereum Virtual Machine (EVM) [10]. Ethereum project is regarded as one of the most prominent platforms for smart contract as its EVM has been adopted by several other platforms such as Hyperledger Burrow and Counterparty [11, 12]. Other platforms are often compared from their compatibility with EVM, as described in TABLE I.

There are platforms that utilize Bitcoin network, for instance Counterparty [9, 13] and Rootstock [14]. Counterparty embeds the transactions data into Bitcoin network. A transaction is encoded in such a way that a Counterparty node can construct its own ledger based on the transactions that has been recorded in Bitcoin network. It is possible to run EVM bytecode smart contract on this platform. Rootstock also utilizes Bitcoin network, but runs on a sidechain. It has its own blockchain, enabling higher throughput, but is secured by the Bitcoin network. This platform implements RVM, an EVM-compatible virtual machine.

Unlike the current implementation of Bitcoin or Ethereum, Stellar Smart Contract does not use proof-of-work algorithm to reach consensus, instead, it uses an algorithm called Stellar Consensus Protocol that is built based on the concept of federated Byzantine agreement [9, 15]. The approach to implement smart contract is *transaction composition* where several transactions are connected and executed by various constraints such as multisignature, batching/atomicy, sequence, and time bound [16].

Hyperledger Burrow [9, 11] is aimed to be a permissioned blockchain platform and utilizes EVM-based architecture to run smart contract. A permissioned blockchain may limit who can participate to the network, therefore a sybil-attack might not be a problem in this kind of network, but a consensus protocol is still needed to provide a canonical state of the blockchain. It is originally contributed by Monax, hosted by Linux Foundation, and co-sponsored by Intel.

TABLE I. VARIOUS PLATFORMS OF BLOCKCHAIN-BASED SMART CONTRACT

| No | Platforms | Contract Language | EVM Compatible |
|---|---|---|---|
| 1 | Bitcoin | Ivy-lang | No |
| 2 | Ethereum | Solidity, Viper | Yes |
| 3 | Counterparty | Solidity | Yes |
| 4 | Rootstock | Solidity | Yes |
| 5 | Stellar Smart Contract | - | No |
| 6 | Hyperledger Burrow | Solidity | Yes |
| 7 | Cardano | Plutus | No |
| 8 | Corda | Java, Kotlin | No |

The next platform is Corda [17], which is aimed for a permissioned blockchain network. Corda uses JVM to run smart contract; therefore, Java based programming language such as Kotlin and Java can be used to code smart contract on it. Corda can also support legal prose to be run as a smart contract.

The most academic-oriented smart contract platform is probably Cardano, as its consensus protocol, Ouroboros, is peer reviewed for Crypto 2017 [18]. Cardano uses IELE architecture to execute its smart contract [19] and Plutus as a strictly typed functional language [20].

IV. SMART CONTRACT BASED APPLICATIONS

There are several applications that have been built on top of the Ethereum smart contract platform, as Ethereum is a Turing-complete platform that allows users to make application on top of it. Due to space limitation, in this section, we only explain two specific examples of applications: tokenization of real-world asset and virtual assets.

An example of blockchain application that tokenizes real world asset is ATLANT, which is a real-estate platform based on blockchain [21]. ATLANT platform pursues a service for individual package of real estate into tokens and PTO (Property Token Offering), then lists them on exchanges based on Ethereum smart contracts. In this system, real-estate unit can be owned by more than one users, because tokenization is able to split real-estate unit into smaller units, which are called tokens.

Another example for blockchain application that tokenizes real world asset is Digix [22]. Claimed as a gold standard in crypto-assets, Digix provides a transparent platform for the tokenization of physical assets by relying in its system on multiple independent participants. This concept was inspired by the nature of money itself in history, which were backed by gold.

For virtual assets, there are a lot of virtual coins and tokens that are already build on top of Ethereum, one of them is Decentraland [23]. Decentraland is a virtual reality platform built on the top of the Ethereum blockchain. Contents and applications can be created, experienced and monetized by the users. The community owns the land in Decentraland as they create it and make them have full control over their creations. The ownership of virtual land can be claimed by Users on a blockchain-based ledger of parcels. The content in the portion of land is fully under control of the land owners, that are identified by a set of Cartesian coordinates (x,y).

V. TOOLS, PROPOSALS, AND BEST PRACTICES

Current academic discussion on this topic are mostly specific on EVM-based smart contract. We could not find any literature that is specific to another platform such as Cardano, Corda, or Stellar Smart Contract. Therefore, in this section we will only focused on Ethereum and EVM-based system. Hopefully, the insights gained in this section could also be applied in a more general context.

Developing a safe Ethereum smart contract program is challenging because of the nature of its immutability, early ecosystem development, and a high incentive to be hacked as it can store economic value [24]. There are already at least 2 cases where bugs on smart contract costed users a huge amount of money. The first was DAO (Decentralised Autonomous Organisation) attack in 2016 that costed about $60 million stolen by the hacker [25]. The second one was Parity wallet bug which froze $146 million being inaccessible [26]. Therefore, to build a safe smart contract is very essential. The security of a smart contract needs to be built properly in order for it to be considered safe to use.

A systematic study conducted by Alharby and Moorsel [27] identified at least four classes regarding smart contract issues: codifying, security, privacy, and performance along with the proposed solutions for each issues. We will revisit each class of these issues and see whether there exists either a new problem or a new solution for the issue, or a solution that is overlooked or excluded by the author.

Pettersson and Edström [28] addresses the problem of codifying issues and proposed a solution by using a dependent and polymorphic types for safer development of smart contract. The result is that dependent effect on functional programming can encode very detailed properties of smart contract behavior, thus reduces the risk of errors and need for testing. Although a proof-of-concept software has been developed, it is not yet ready for production use due to large size of compiled code and incomplete implementation of the theory. The authors conclude that it is not clear whether functional programming is beneficial to develop smart contract. Therefore, a further study is needed to determine suitable paradigm for programming language of smart contract.

In [29], the problem of codifying issues is also addressed. The author proposes an intermediate level language between a high-level language (e.g. Solidity) to the Ethereum Virtual Machine. The tool called Scilla, its main feature is a clean separation between the communication aspect of smart contract that allows a rich interaction patterns and a programmable component which enables principled semantics and easy compatible to formal verification.

As for the security issues, there are at least 11 known vulnerabilities in EVM-based Smart Contract: reentrancy, unchecked send call, failed send, integer overflow/underflow, transaction state dependence, absence of logic, incorrect logic, logically correct but unfair, block state dependence, transaction order dependence, and trace vulnerabilities [24, 30]. Symbolic analysis tool such as Oyente, Manticore, Mythrill, Securify, SmartCheck and KEVM are often used to detect the vulnerabilities [24, 31, 32, 33, 34, 35, 36].

A new tool called Zeus is proposed in [30]. Zeus is a symbolic analysis tool to verify the correctness and validate the fairness of smart contract. It is claimed to significantly outperform Oyente, but a source code and correctness specifications must be provided. Concurrently, other tool called Maian is proposed in [24]. The tool is able to detect Parity bug by specifying and reason trace properties by defining systematic characterization of trace vulnerabilities security issue.

With regard to codifying and security, ConsenSys publishes a documentation focusing on covering the best practices on Ethereum Smart Contract Security [37]. In the document, smart contract development is very different from standard software engineering approach, as the cost of failure can be high. Hence, smart contract development must be well planned, similar to hardware engineering.

As also been stated in Alharby and Moorsel's study, on the matter of privacy issues, Hawk is proposed in [38]. It is a model of cryptography and privacy-preserving smart contract for blockchain. Furthermore, Town Crier [39] is also proposed as a model of authentication in data for smart contracts.

Several proposals related to performance issues have been proposed. In [40, 41], to make Ethereum scalable, sharding and state channel (a.k.a. Plasma) [42] could be applied. Sharding is an approach to divide the blockchain into several chains that are working in parallel but still can be viewed as one chain, while Plasma is a solution that allows the creation of a blockchain on top of another blockchain. However, both approaches are still in development, and several problems need to be solved before they are ready for implementation [40, 41, 42].

## VI. DISCUSSION

This paper compiles several recent literature regarding blockchain-based smart contract technology. We have discussed several smart contract platforms based on Bartoletti and Pompianu's survey [9] and expand it upon Igor Korsakov's list [8]. A more comprehensive study on the matter of differences between various blockchain-based smart contract platforms could be addressed in future studies.

The current blockchain-based smart contract platforms is divided into two main architectures: EVM-compatible platform and non EVM-compatible platform. As stated by Alharby and van Moorsel in their systematic study [27], almost all research on smart contract is done on Ethereum platform. While it might be also applicable to EVM-compatible platforms, other non EVM-compatible platforms also has their own uniqueness propositions that need to be investigated if some smart contract are built upon it, especially issues related to security problems. Furthermore, Alharby and van Moorsel also identified several other research gaps such as the lack of study on scalability and performance issue, lack of application on smart contract in academic field, lack of research to tackle criminal activities in smart contract, and the lack of peer-reviewed paper that are discussing the topic of smart contract.

Moreover, in the investigation on the use of functional programming to build smart contract by Pettersson and Edström [28], it is not clear whether functional programming is beneficial in smart contract development. While dependent and polymorphic types might be able to encode very detailed properties of smart contract behaviour, the large size of the compiled program and the incomplete implementation of the theory.causes the difficulty in using functional programming for the purpose. Thus, research to determine the most suitable paradigm for smart contract programming is yet to be conducted in future research.

With regard to the lack of application on smart contract in academic field, we build a preliminary framework for the application of smart contract in academic field, especially in Indonesia.

A possible implementation of a blockchain-based smart contract is to establish a consortium network of various higher education institutes and universities for digital academic diploma system in Indonesia.

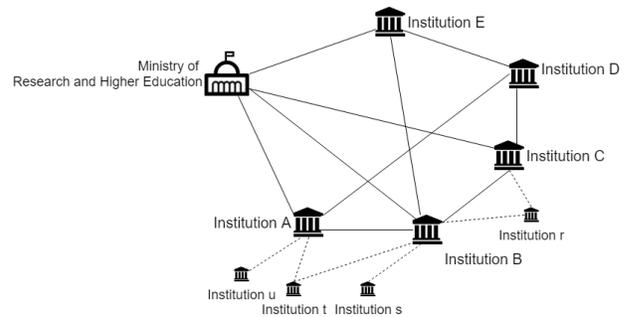

Fig. 2. Architecture of a Consortium Network for Academic Diploma in Indonesia

The system will utilize a private blockchain network such as Hyperledger Burrow. The network can be configured such that it will have a set of institutions as network validators, a set of smaller institutions (which might not have as much resources as validator institutions) as piggyback nodes, and the Ministry of Research and Higher Education as manager node to manage the validator nodes and piggyback nodes members.

When an institution will issue a diploma, a JSON file with fields containing information of the diploma will be created and a hash file of the JSON diploma is generated. Each fields of data of the diploma then is appended with the JSON's hash and then the appended string is hashed. The hashes then can be published to the network by calling a function of a smart contract from an institution's node (hence, the transaction must be signed by the institution). Each JSON file is kept by its respective diploma owners. This way, the data authenticity can be preserved while not exposing the raw data.

When a party X wants to verify a field from a diploma, X must have several information i.e. JSON's hash, field name, field data and issuer's address. The field data is appended to the JSON's hash and checked whether it is exist on the blockchain and signed by the issuer.

## VII. Conclusion

Blockchain technology is invented alongside Bitcoin, a peer-to-peer digital cash, out of the need to create a single-point-of-failure resistant decentralized system that can prevent double spending problem and sybil-attack. It turns out that it can be utilized in several fields, including the implementation of self-enforcing smart contract system. There are several platforms that provides this system such as Ethereum, Counterparty, Hyperledger, and many others.

Since blockchain-based smart contract technology is a growing field, wide opportunities are available for more research for future work. More study on scalability and performance issue is needed. With regard to security, there is only a few of research tackle criminal activities in smart contract. In terms of applications, more tools should be built for blockchain-based smart contract technology, especially on the platform that are not EVM-based. Exploration to find suitable paradigm for programming language of smart contract is required. We also propose a preliminary framewrok for the application of blockchain-based smart contract in the acedemic field.

Lastly, more comprehensive study on blockchain-based smart contract technology is yet to be conducted, especially to discover the advantages as well as disadvantages of each platform. This kind of study is beneficial for decision makers in decision making process to select the most suitable blockchain-based smart contract application to be adopted in the environment.